\documentclass{jpaper}

\usepackage[nocompress]{cite}
\usepackage{algorithmic}
\usepackage{array}

\makeatletter
\let\MYcaption\@makecaption
\makeatother

\usepackage[font=footnotesize]{subcaption}

\makeatletter
\let\@makecaption\MYcaption
\makeatother

\usepackage{fixltx2e}
\usepackage{dblfloatfix}
\usepackage[nolessnomore, italic]{mathastext}
\usepackage[T1]{fontenc}
\usepackage[usenames,dvipsnames,svgnames,table]{xcolor}
\usepackage[normalem]{ulem}
\usepackage{enumitem}
\usepackage{setspace}
\usepackage{indentfirst}
\usepackage{footmisc}
\usepackage{fancyhdr}
\usepackage{authblk}
\usepackage{url}
\usepackage[binary-units=true]{siunitx}
\usepackage[us,12hr]{datetime}
\usepackage[keeplastbox]{flushend}
\usepackage[hidelinks]{hyperref}

\newif\ifcameraready
\camerareadytrue

\newcommand{\versionnum}[0]{7}

\ifcameraready
\else
\fi

\ifcameraready
  \newcommand{\todo}[1][]{}
  \newcommand{\chI}[0]{}
  \newcommand{\chII}[0]{}
  \newcommand{\chIII}[0]{}
  \newcommand{\chIV}[0]{}
  \newcommand{\ch}[0]{}
\else
  \newcommand{\todo}[1][]{\textbf{\fcolorbox{black}{red}{\color{white}{TODO}}} \underline{$\overline{\hbox{\emph{#1}}}$}}
  \newcommand{\chI}[0]{}
  \newcommand{\chII}[0]{}
  \newcommand{\chIII}[0]{}
  \newcommand{\chIV}[1]{{\textcolor{MidnightBlue}{#1}}}
  \newcommand{\ch}[1]{{\textcolor{BrickRed}{#1}}}
\fi

\begin{document}
%
\title{Guest Editor Introduction:\vspace{-1.5pt}\\ \chIII{Recent Advances in Overcoming Bottlenecks in Memory Systems\vspace{-1.5pt}\\ and Managing Memory Resources in GPU Systems}}


\author{%
{Onur Mutlu$^{1,2}$}%
\qquad%
{Saugata Ghose$^{2}$}%
\qquad%
{Rachata Ausavarungnirun$^{2}$}%
}

\affil{\em%
$^{1}$ETH Z{\"u}rich%
\qquad%
$^{2}$Carnegie Mellon University%
}

\maketitle


\chI{Memory and storage systems are a fundamental system
performance, energy, and reliability bottleneck in modern 
systems\chI{~\cite{mutlu.imw13, mutlu.superfri14, mutlu.date17,
cai.procieee17, cai.procieee.arxiv17, cai.bookchapter.arxiv17}}.
This bottleneck is becoming increasingly severe due to 
(1)~the \chII{very} limited latency reductions in memory and storage devices over
the last several years;
(2)~aggressive manufacturing process technology scaling and \chII{other
techniques to improve memory density,}
such as multi-level cell technology, which 
\chII{increase} the \chII{storage capacity} of these devices, but introduce 
more raw bit errors and increase manufacturing process variation;
(3)~limited pin counts in chip packages, which \chII{prevent} system designers
from adding more and/or wider buses to increase bandwidth;
(4)~\chII{overwhelmingly} data-intensive applications, which 
\chII{require high-bandwidth access to very large amounts of data;} and
(5)~the increasing fraction of overall system energy consumed by memory
systems and data movement.
To make matters worse, it is becoming increasingly difficult to continue
scaling these devices to smaller process \chII{technology nodes}, and \chII{even though}
alternative \chII{emerging} memory and storage technologies \chII{can potentially} 
\chIII{alleviate} some of the 
\chII{shortcomings of existing memory and storage technologies, they also}
introduce \emph{new} \chII{shortcomings that were previously absent}.  
Therefore, there is a pressing need to
comprehensively understand and mitigate these bottlenecks in both
existing and emerging \chII{memory and storage} systems and technologies.}

This issue features extended summaries \chII{and retrospectives} of some of the \chII{recent research}
done by our research group, SAFARI~\cite{safari.website, safari.github},
on \chI{\chIII{(1)~}various critical problems in memory systems 
and \chIII{(2)~}how memory \chII{system} bottlenecks affect 
\chII{graphics processing unit (GPU)} systems.
As more applications share a single system, operations from each application 
can \chIII{contend} with each other at various shared components within the system.
If left unmitigated, \chIII{such contention} can undermine
many of the benefits of parallelism, by slowing down each application
\chII{or thread of execution\chIII{~\cite{mutlu.isca08, moscibroda.usenixsecurity07,
mutlu.micro07, mutlu.superfri14, mutlu.imw13}}}.
\chIII{The compound effect of contention, high memory latency and access overheads,
as well as inefficient management of resources, greatly degrades
performance, quality-of-service (QoS), and energy efficiency.}
The ten works featured in this issue study several aspects of 
(1)~inter-application interference in multicore systems, heterogeneous systems, and
GPUs; 
(2)~the growing overheads and expenses associated with growing memory densities
and latencies; and
(3)~\chII{performance,} programmability, and portability issues in modern GPUs,
\chII{especially those related to memory \chIII{system} resources}.

\chIII{These works rely on real system characterizations and simulation to develop
a rigorous understanding of the interference and bottlenecks, and to provide
solutions.  Our analyses have shown key scaling \chIV{and performance} bottlenecks, proposed new
solutions, and have inspired the research community to develop further
investigations (e.g., \chIV{on} interference and fairness in main 
memory~\cite{subramanian.hpca13, subramanian.micro15, subramanian.iccd14,
subramanian.tpds16},
subarray-level parallelism~\cite{kim.isca12, chang.hpca14}, low-cost memory reliability~\cite{luo.dsn14},
hybrid memory management\chIV{~\cite{yoon.iccd12, li.cluster17, meza.cal12, 
ren.micro15, meza.weed13}}).
In order to \ch{aid future} research, we have released our flexible and
\chIV{extensible} memory system simulator, Ramulator, as open-source
software~\cite{kim.cal15, ramulator.github}, and have released open-source simulators
that accurately model memory interference in multicore 
systems\chIV{~\cite{asmsim.github, memschedsim.github}}
and memory resource bottlenecks in GPU systems~\cite{medic.github, mosaic.github}.}

In each work \chIII{that is featured in this issue}, based on \chIII{our} rigorous studies
\chIII{and analyses}, we propose novel solutions that mitigate
many of these problems.}
\chII{We examine GPUs as a special example because they enable massively parallel
processing on a single chip and, as a result, are limited greatly by the 
bottlenecks in the memory system.}
%
For \chII{each of the} works presented \chI{in this special issue}, 
\chII{its corresponding article examines the} work's significance in the
context of modern computer systems, and \chII{discusses} several \chIII{new} research
questions \chIII{and directions} that each work motivates.

We start with \chI{three of our works} that manage interference and contention in 
\chII{main memory}. 
\chII{When multiple applications (or multiple threads of \chIII{one or more} applications) 
concurrently issue memory requests, these requests often contend with each
other \chIII{in the main memory system}, increasing the average memory access latency
\chIII{and reducing per-application or per-thread parallelism}.  This contention becomes
especially problematic when a highly-memory-intensive application issues many
more requests than other applications, causing requests from the
other applications to unfairly wait for very long times as the
memory system \chIII{takes time to service} all of the requests from the highly-memory-intensive
application.  To mitigate the interference that \chIII{each}
application \chIII{induces} on the other applications, memory systems must adopt \chIII{new}
mechanisms 
to regulate the 
available memory \chIII{bandwidth} among all applications and/or reduce the amount of
memory-level contention.}
\chIII{Doing so can enable systems that are higher performance, more
predictable, and more energy efficient at the same time.  The first three works
featured in this issue enable new mechanisms to more efficiently manage
\chIV{interference and} contention in main memory.}

\chI{The first paper in the issue~\cite{predictable-safarij} describes} Memory Interference-induced Slowdown Estimation (MISE),
which originally appeared in HPCA 2013~\cite{subramanian.hpca13}.
\chI{This work \chI{(1)~}develops} a model \chII{called MISE, which} predicts the impact of interference in DRAM on the
overall system performance; and \chII{(2)~}\chI{uses} this model to design \chII{new} memory schedulers that
improve fairness \chIII{and QoS} among concurrently-executing applications.
\chII{The work finds that \chIII{various} MISE-based memory schedulers 
\chIII{can (1)~provide predictable performance to designated applications and 
(2)~}significantly improve the overall system throughput.}

\chI{The second paper in the issue~\cite{sms-safarij} describes} Staged Memory Scheduling, 
which originally appeared in ISCA 2012~\cite{ausavarungnirun.isca12}.
\chI{This work \chIII{analyzes}} the high impact of interference between the CPU and GPU in a
heterogeneous system (e.g., a system-on-chip),
\chIII{showing that the GPU can overwhelm CPU performance and sometimes
vice versa}.  \chII{Based on this finding, the work}
\chI{develops} a \chII{new} memory
controller that provides fair memory access for both CPU and GPU applications,
\chII{improving the performance of CPU applications without affecting the
throughput of GPU applications}.

\chI{The third paper in the issue~\cite{salp-safarij} describes} \chII{Subarray-Level Parallelism} (SALP), 
which originally appeared in ISCA 2012~\cite{kim.isca12}.
\chI{This work exploits} the subarrays (i.e., sub-banking) in DRAM 
architectures to greatly increase the amount of \chIII{memory} parallelism available
to applications.  SALP proposes three \chII{new} mechanisms to expose the subarrays to the
memory controller \chIII{at low cost}, improving row locality and reducing the number of
high-latency \chII{bank} conflicts that occur
when multiple \chIII{requests} access the same memory bank.
\chII{The reduced \chIII{bank conflicts and the improved row locality
significantly improve} overall system performance \chIV{and reduce energy
consumption}.}

Next, we \chI{look at several of our works that}
address the growing overheads and expenses associated with
growing \chII{main} memory densities \chI{and latencies}.  
As systems execute more applications in parallel, and
as applications process larger amounts of data, DRAM manufacturers have relied on
\chIII{aggressive technology}
scaling to increase the density of each DRAM device.  Unfortunately, \chIII{such} scaling has
introduced a number of \chII{key} challenges\chIV{~\cite{mutlu.imw13, 
mutlu.superfri14, mutlu.date17}}, which we \chII{methodically} address
\chIII{in the next four works}.

\chI{Our fourth paper in the issue~\cite{dsarp-safarij} describes} DSARP, 
which originally appeared in HPCA 2014~\cite{chang.hpca14}.
\chI{This work explores} how increasing \chIII{memory density} will cause 
DRAM refresh operations to \chIV{become a bigger performance bottleneck}, 
preventing the DRAM from \chII{effectively} servicing
outstanding memory requests \chII{with low latency}.  \chI{The work proposes} new memory controller
policies that almost completely eliminate the \chIII{performance overhead of DRAM refresh}
by performing refresh operations in the background \chIV{via low-cost changes to
the DRAM architecture and the memory controller}.

\chI{Our fifth paper in the issue~\cite{chargecache-safarij} describes} ChargeCache, 
which originally appeared in HPCA 2016~\cite{hassan.hpca16}.
\chI{This work finds} that many applications must reopen memory rows soon after they are closed
because of interference \chIII{(i.e., bank conflicts)}, incurring a high access latency.
ChargeCache \chIII{is a new mechanism that} takes advantage of the high charge 
held within a recently-closed row
to reduce the access latency to such a row \chIV{when it is accessed again soon
in the future.  The work shows that ChargeCache significantly} \chIII{improves the overall 
system performance and energy consumption}.

\chI{Our sixth paper in the issue~\cite{hrm-safarij} describes} heterogeneous-reliability memory (HRM), 
which originally appeared in DSN 2014~\cite{luo.dsn14}.
\chI{This work demonstrates} on real machines that many data center applications can tolerate 
errors in large regions of \chIV{their memory address spaces} without affecting 
correctness.  \chI{The work uses} this 
observation to lower the cost of memory subsystems for data centers, by 
\chIII{introducing a new memory system framework, HRM, where 
\chIV{the memory system consists of different modules with different types and
amounts of error correction/detection capabilities.}
By employing many DRAM modules without error correction and
intelligently mapping error-tolerant memory regions to these modules
\chIV{and error-vulnerable memory regions to DRAM modules with error correction},
HRM significantly reduces the cost of a data center system, while still
providing high overall reliability \chIV{and availability}.}

\chI{Our seventh paper in the issue~\cite{rbla-safarij} describes} row buffer locality aware (RBLA) caching, 
which originally appeared in ICCD 2012~\cite{yoon.iccd12}.
\chI{This work proposes} a new \chIV{technique} to manage data placement in hybrid memory systems, which combine
conventional DRAM with emerging memory technologies to provide the benefits
of both in a scalable yet cost-effective manner.
\chIV{Exploiting the key observation that row buffer hits are of the same cost in
both DRAM and emerging memory technologies,}
\chIII{RBLA avoids migrating data from the emerging memory to conventional DRAM
(and vice versa)
when the migration would not yield a significant benefit,
\chIV{thereby preserving the precious DRAM space for data that really benefits 
from the low access latency of DRAM arrays.}
\chIV{The work shows that RBLA} improves both
system performance and energy consumption as a result.}

Finally, we examine how \ch{to} manage \chI{memory} \chIII{resources} within GPUs.  For many
general-purpose GPU (GPGPU) applications, programmers are responsible for
\chIII{explicitly managing all memory resources, including registers},
by \chIII{specifying in programs how much each application should get of each resource}.
Our solutions \chIII{automatically manage}
these resources in \chIV{both} hardware \chIII{and software, \chIV{and sometimes
cooperatively between the hardware and software,} transparently to the programmer}.
\chIII{The solutions lift the burden of resource management from the programmer, 
and improve the performance and efficiency of GPGPU applications.}

\chI{Our eighth paper in the issue~\cite{zorua-safarij} describes} Zorua, 
which originally appeared in MICRO 2016~\cite{vijaykumar.micro16}.
\chIII{Current GPU systems require programmers to discover and explicitly specify
the quantities of each resource that are assigned to a thread, in order to avoid
significant performance penalties.}
\chI{This work proposes} a \chIII{new} resource virtualization mechanism for GPGPU applications,
\ch{called Zorua,}
\chIII{which can assign resources to each thread dynamically at runtime based on
the thread's needs and the available resources in the GPU, \chIV{with only
annotations provided by the compiler}.
\ch{With its effective resource virtualization, Zorua improves
(1)~programmability, by removing the existing burden on programmers to
tune the thread resource allocation;
(2)~portability, by removing the need to retune the resource
allocation when an application tuned for one
GPU architecture is executed on a different GPU architecture; and
(3)~performance, by ensuring the careful allocation and oversubscription of
resources to best utilize the hardware.}}

\chI{Our ninth paper in the issue~\cite{medic-safarij} describes} Memory Divergence Correction (MeDiC), 
which originally appeared in PACT 2015~\cite{ausavarungnirun.pact15},
\chI{This work finds} that different warps (i.e., groups of threads that execute in lockstep)
exhibit different levels of memory divergence, where some, but not all, threads
stall on long-latency memory accesses, which prevents forward progress for all threads in the warp.
MeDiC consists of three \chIII{new} mechanisms that work together to optimize 
cache and memory \chIV{resource management} 
in a GPU, based on the divergence behavior of the warps belonging to an application.
\chIII{These three mechanisms provide \chIV{significant performance improvements} for
GPGPU applications.}

\chI{Our tenth paper in the issue~\cite{mosaic-safarij} describes} Mosaic, 
which originally appeared in MICRO 2017~\cite{ausavarungnirun.micro17}.
\ch{In contemporary GPUs, limited resources for memory virtualization
can cause a single operation (e.g., an address translation that misses in the GPU's
translation lookaside buffer) to often stall hundreds of threads for long latencies, 
leading to significant underutilization of the GPU.}
\ch{The memory virtualization bottleneck can be alleviated by changing the page
size, but a major hurdle to this is the}
\chIV{key trade-off between \ch{two costly operations:}
demand paging (which benefits from small page sizes) and address translation
(which benefits from large page sizes).}
\chI{This work proposes} a \chIII{new} \ch{hardware mechanism that
takes advantage of GPGPU memory access patterns to enable the
efficient support of}
multiple page sizes transparently to the programmer.
\chIII{\ch{By efficiently supporting multiple page sizes, Mosaic alleviates the
high contention for memory virtualization resources, which in turn significantly}
improves the performance of GPGPU applications.}

\ch{Throughout all of these works, we
(1)~identify various points of interference, contention, and resource 
bottlenecks in memory systems and GPUs; and
(2)~appropriately modify the systems to mitigate these issues at low cost and
low overhead.
These works improve the performance, fairness, energy consumption, and/or
programmability of a system, and often improve scalability as more applications
execute concurrently on the system.}
\ch{Even though the works presented are described in the context of DRAM,
the dominant memory technology of today,
we believe many of the basic ideas and concepts can be applied or adapted to
emerging memory technologies~\cite{meza.weed13},
e.g., phase-change memory~\cite{lee.isca09, qureshi.isca09, wong.procieee10, 
lee.ieeemicro10, zhou.isca09, lee.cacm10, yoon.taco14},
STT-MRAM~\cite{naeimi.itj13, kultursay.ispass13, guo.isca10},
and memristors/RRAM~\cite{wong.procieee12, chua.tct71, strukov.nature08}.}
We hope that \chIII{the works featured in this special issue} inspire readers to explore other sources of 
\chIII{interference, \ch{contention,} performance, \ch{and programmability} issues} in
modern systems, and
to develop new solutions that can enable fair, high-performance,
\chIII{energy-efficient} systems for the future.

\section*{Acknowledgments}

The works featured in this issue, along with our related works that
we reference in each \chIII{featured work}, are a
result of the research done together with many students and
collaborators over the course of the past 10+ years, whose
contributions we acknowledge.
In particular, we acknowledge and appreciate the dedicated effort 
of current and former students and postdocs in our research group,
\chI{SAFARI~\cite{safari.website, safari.github}},
who contributed to the \chI{ten} featured works, 
including
Kevin Chang,
Rachael Harding,
Hasan Hassan,
Kevin Hsieh,
Ben Jaiyen,
Samira Khan, 
Yoongu Kim, 
Donghyuk Lee, 
Yixin Luo, 
Justin Meza, 
Gennady Pekhimenko, 
Vivek Seshadri, 
Ashish Shrestha,
Lavanya Subramanian, 
Nandita Vijaykumar, and
HanBin Yoon.

Aside from our featured
works and other referenced papers from our group, where a wealth
of information on modern memory and storage systems can be found, at least four
Ph.D.\ dissertations have shaped the works that we feature \chI{in this special issue}:
\begin{itemize}[leftmargin=1.6em]
\item Lavanya Subramanian's thesis entitled ``Providing
High and Controllable Performance in Multicore Systems Through Shared
Resource Management''~\cite{subramanian.thesis15},

\item Yoongu Kim's thesis
entitled ``Architectural Techniques to Enhance DRAM
Scaling''~\cite{kim.thesis15},

\item Kevin Chang's thesis entitled ``Understanding and
Improving the Latency of DRAM-Based Memory
Systems''~\cite{chang.thesis17}, and

\item Rachata Ausavarungnirun's
thesis entitled ``Techniques for Shared Resource Management in Systems
with Throughput Processors''~\cite{ausavarungnirun.thesis17}.
\end{itemize}

We also acknowledge various funding agencies (the National Science
Foundation, the Semiconductor Research Corporation, the Intel Science and
Technology Center on Cloud Computing, CyLab, the CMU Data Storage
Systems Center, and the NIH) and
industrial partners (AMD, Facebook, Google, HP Labs, Huawei, IBM,
Intel, Microsoft, NVIDIA, Oracle, Qualcomm, Rambus, Samsung, Seagate,
VMware), and ETH Z{\"u}rich, who have supported the featured works in
this issue and other related work \chIII{in our research group} generously 
over the years.

{
\bibliographystyle{IEEEtranS}
\bibliography{refs}

\begin{thebibliography}{10}
\providecommand{\url}[1]{#1}
\csname url@samestyle\endcsname
\providecommand{\newblock}{\relax}
\providecommand{\bibinfo}[2]{#2}
\providecommand{\BIBentrySTDinterwordspacing}{\spaceskip=0pt\relax}
\providecommand{\BIBentryALTinterwordstretchfactor}{4}
\providecommand{\BIBentryALTinterwordspacing}{\spaceskip=\fontdimen2\font plus
\BIBentryALTinterwordstretchfactor\fontdimen3\font minus
  \fontdimen4\font\relax}
\providecommand{\BIBforeignlanguage}[2]{{%
\expandafter\ifx\csname l@#1\endcsname\relax
\typeout{** WARNING: IEEEtranS.bst: No hyphenation pattern has been}%
\typeout{** loaded for the language `#1'. Using the pattern for}%
\typeout{** the default language instead.}%
\else
\language=\csname l@#1\endcsname
\fi
#2}}
\providecommand{\BIBdecl}{\relax}
\BIBdecl

\bibitem{mosaic-safarij}
R.~Ausavarungnirun, J.~Landgraf, V.~Miller, S.~Ghose, J.~Gandhi, C.~J.
  Rossbach, and O.~Mutlu, ``{Mosaic: An Application-Transparent
  Hardware-Software Cooperative Memory Manager for GPUs},''
  \url{https://arxiv.org/abs/1804.11265}.

\bibitem{ausavarungnirun.micro17}
R.~Ausavarungnirun, J.~Landgraf, V.~Miller, S.~Ghose, J.~Gandhi, C.~J.
  Rossbach, and O.~Mutlu, ``{Mosaic: A GPU Memory Manager with
  Application-Transparent Support for Multiple Page Sizes},'' in \emph{MICRO},
  2017.

\bibitem{ausavarungnirun.thesis17}
R.~Ausavarungnirun, ``{Techniques for Shared Resource Management in Systems
  with Throughput Processors},'' Ph.D. dissertation, Carnegie Mellon Univ.,
  2017.

\bibitem{sms-safarij}
R.~Ausavarungnirun, K.~K. Chang, L.~Subramanian, G.~H. Loh, and O.~Mutlu,
  ``{High-Performance and Energy-Effcient Memory Scheduler Design for
  Heterogeneous Systems},'' \url{https://arxiv.org/abs/1804.11043}.

\bibitem{ausavarungnirun.isca12}
R.~Ausavarungnirun, K.~K. Chang, L.~Subramanian, G.~H. Loh, and O.~Mutlu,
  ``{Staged Memory Scheduling: Achieving High Performance and Scalability in
  Heterogeneous Systems},'' in \emph{ISCA}, 2012.

\bibitem{medic-safarij}
R.~Ausavarungnirun, S.~Ghose, O.~Kayıran, G.~H. Loh, C.~R. Das, M.~T.
  Kandemir, and O.~Mutlu, ``{Holistic Management of the GPGPU Memory Hierarchy
  to Manage Warp-level Latency Tolerance},''
  \url{https://arxiv.org/abs/1804.11038}.

\bibitem{ausavarungnirun.pact15}
R.~Ausavarungnirun, S.~Ghose, O.~Kayıran, G.~H. Loh, C.~R. Das, M.~T.
  Kandemir, and O.~Mutlu, ``{Exploiting Inter-Warp Heterogeneity to Improve
  GPGPU Performance},'' in \emph{PACT}, 2015.

\bibitem{cai.procieee17}
Y.~Cai, S.~Ghose, E.~F. Haratsch, Y.~Luo, and O.~Mutlu, ``{Error
  Characterization, Mitigation, and Recovery in Flash-Memory-Based Solid-State
  Drives},'' \emph{Proc. IEEE}, 2017.

\bibitem{cai.procieee.arxiv17}
Y.~Cai, S.~Ghose, E.~F. Haratsch, Y.~Luo, and O.~Mutlu, ``{Error
  Characterization, Mitigation, and Recovery in Flash Memory Based Solid-State
  Drives},'' arXiv:1706.08642 [cs.AR], 2017.

\bibitem{cai.bookchapter.arxiv17}
Y.~Cai, S.~Ghose, E.~F. Haratsch, Y.~Luo, and O.~Mutlu, ``{Errors in
  Flash-Memory-Based Solid-State Drives: Analysis, Mitigation, and Recovery},''
  arXiv:1711.11427 [cs.AR], 2017.

\bibitem{dsarp-safarij}
K.~K. Chang, D.~Lee, Z.~Chishti, A.~R. Alameldeen, C.~Wilkerson, Y.~Kim, and
  O.~Mutlu, ``{Reducing DRAM Refresh Overheads with Refresh-Access
  Parallelism},'' \url{https://arxiv.org/abs/1805.01289}.

\bibitem{chang.hpca14}
K.~K. Chang, D.~Lee, Z.~Chishti, A.~R. Alameldeen, C.~Wilkerson, Y.~Kim, and
  O.~Mutlu, ``{Improving DRAM Performance by Parallelizing Refreshes with
  Accesses},'' in \emph{HPCA}, 2014.

\bibitem{chang.thesis17}
K.~K. Chang, ``{Understanding and Improving the Latency of DRAM-Based Memory
  Systems},'' Ph.D. dissertation, Carnegie Mellon Univ., 2017.

\bibitem{chua.tct71}
L.~Chua, ``{Memristor---The Missing Circuit Element},'' \emph{TCT}, 1971.

\bibitem{guo.isca10}
X.~Guo, E.~Ipek, and T.~Soyata, ``{Resistive Computation: Avoiding the Power
  Wall with Low-Leakage, STT-MRAM Based Computing},'' in \emph{ISCA}, 2010.

\bibitem{chargecache-safarij}
H.~Hassan, G.~Pekhimenko, N.~Vijaykumar, V.~Seshadri, D.~Lee, O.~Ergin, and
  O.~Mutlu, ``{Exploiting Row-Level Temporal Locality in DRAM to Reduce the
  Memory Access Latency},'' \url{https://arxiv.org/abs/1805.03969}.

\bibitem{hassan.hpca16}
H.~Hassan, G.~Pekhimenko, N.~Vijaykumar, V.~Seshadri, D.~Lee, O.~Ergin, and
  O.~Mutlu, ``{ChargeCache: Reducing DRAM Latency by Exploiting Row Access
  Locality},'' in \emph{HPCA}, 2016.

\bibitem{salp-safarij}
Y.~Kim, V.~Seshadri, D.~Lee, J.~Liu, and O.~Mutlu, ``{Exploiting the DRAM
  Microarchitecture to Increase Memory-Level Parallelism},''
  \url{https://arxiv.org/abs/1805.01966}.

\bibitem{kim.isca12}
Y.~Kim, V.~Seshadri, D.~Lee, J.~Liu, and O.~Mutlu, ``{A Case for Exploiting
  Subarray-Level Parallelism (SALP) in DRAM},'' in \emph{ISCA}, 2012.

\bibitem{kim.thesis15}
Y.~Kim, ``{Architectural Techniques to Enhance DRAM Scaling},'' Ph.D.
  dissertation, Carnegie Mellon Univ., 2015.

\bibitem{kim.cal15}
Y.~Kim, W.~Yang, and O.~Mutlu, ``{Ramulator: A Fast and Extensible DRAM
  Simulator},'' \emph{CAL}, 2015.

\bibitem{kultursay.ispass13}
E.~K{\"u}lt{\"u}rsay, M.~Kandemir, A.~Sivasubramaniam, and O.~Mutlu,
  ``{Evaluating STT-RAM as an Energy-Efficient Main Memory Alternative},'' in
  \emph{ISPASS}, 2013.

\bibitem{lee.isca09}
B.~C. Lee, E.~Ipek, O.~Mutlu, and D.~Burger, ``{Architecting Phase Change
  Memory as a Scalable DRAM Alternative},'' in \emph{ISCA}, 2009.

\bibitem{lee.cacm10}
B.~C. Lee, E.~Ipek, O.~Mutlu, and D.~Burger, ``{Phase Change Memory
  Architecture and the Quest for Scalability},'' \emph{CACM}, 2010.

\bibitem{lee.ieeemicro10}
B.~C. Lee, P.~Zhou, J.~Yang, Y.~Zhang, B.~Zhao, E.~Ipek, O.~Mutlu, and
  D.~Burger, ``{Phase-Change Technology and the Future of Main Memory},''
  \emph{IEEE Micro}, 2010.

\bibitem{li.cluster17}
Y.~Li, S.~Ghose, J.~Choi, J.~Sun, H.~Wang, and O.~Mutlu, ``{Utility-Based
  Hybrid Memory Management},'' in \emph{CLUSTER}, 2017.

\bibitem{hrm-safarij}
Y.~Luo, S.~Govindan, B.~Sharma, M.~Santaniello, J.~Meza, A.~Kansal, J.~Liu,
  B.~Khessib, K.~Vaid, and O.~Mutlu, ``{Heterogeneous-Reliability Memory:
  Exploiting Application-Level Memory Error Tolerance`},''
  \url{https://arxiv.org/abs/1602.00729}.

\bibitem{luo.dsn14}
Y.~Luo, S.~Govindan, B.~Sharma, M.~Santaniello, J.~Meza, A.~Kansal, J.~Liu,
  B.~Khessib, K.~Vaid, and O.~Mutlu, ``{Characterizing Application Memory Error
  Vulnerability to Optimize Datacenter Cost via Heterogeneous-Reliability
  Memory},'' in \emph{DSN}, 2014.

\bibitem{meza.weed13}
J.~Meza, Y.~Luo, S.~Khan, J.~Zhao, Y.~Xie, and O.~Mutlu, ``{A Case for
  Efficient Hardware-Software Cooperative Management of Storage and Memory},''
  in \emph{WEED}, 2013.

\bibitem{meza.cal12}
J.~Meza, J.~Chang, H.~Yoon, O.~Mutlu, and P.~Ranganathan, ``{Enabling Efficient
  and Scalable Hybrid Memories Using Fine-Granularity {DRAM} Cache
  Management},'' \emph{CAL}, 2012.

\bibitem{moscibroda.usenixsecurity07}
T.~Moscibroda and O.~Mutlu, ``{Memory Performance Attacks: Denial of Memory
  Service in Multi-Core Systems},'' in \emph{USENIX Security}, 2007.

\bibitem{mutlu.date17}
O.~Mutlu, ``{The RowHammer Problem and Other Issues We May Face as Memory
  Becomes Denser},'' in \emph{DATE}, 2017.

\bibitem{mutlu.micro07}
O.~Mutlu and T.~Moscibroda, ``{Stall-Time Fair Memory Access Scheduling for
  Chip Multiprocessors},'' in \emph{MICRO}, 2007.

\bibitem{mutlu.isca08}
O.~Mutlu and T.~Moscibroda, ``{Parallelism-Aware Batch Scheduling: Enhancing
  Both Performance and Fairness of Shared DRAM Systems},'' in \emph{ISCA},
  2008.

\bibitem{mutlu.imw13}
O.~Mutlu, ``{Memory Scaling: A Systems Architecture Perspective},'' in
  \emph{IMW}, 2013.

\bibitem{mutlu.superfri14}
O.~Mutlu and L.~Subramanian, ``{Research Problems and Opportunities in Memory
  Systems},'' \emph{SUPERFRI}, 2014.

\bibitem{naeimi.itj13}
H.~Naeimi, C.~Augustine, A.~Raychowdhury, S.-L. Lu, and J.~Tschanz, ``{STT-RAM
  Scaling and Retention Failure},'' \emph{Intel Technology Journal}, 2013.

\bibitem{qureshi.isca09}
M.~K. Qureshi, V.~Srinivasan, and J.~A. Rivers, ``{Scalable High Performance
  Main Memory System Using Phase-Change Memory Technology},'' in \emph{ISCA},
  2009.

\bibitem{ren.micro15}
J.~Ren, J.~Zhao, S.~Khan, J.~Choi, Y.~Wu, and O.~Mutlu, ``{ThyNVM: Enabling
  Software-Transparent Crash Consistency in Persistent Memory Systems},'' in
  \emph{MICRO}, 2015.

\bibitem{safari.website}
{SAFARI Research Group}, \url{http://www.ece.cmu.edu/~safari/}.

\bibitem{asmsim.github}
{SAFARI Research Group}, ``{ASMSim -- GitHub Repository},''
  \url{https://github.com/CMU-SAFARI/ASMSim}.

\bibitem{medic.github}
{SAFARI Research Group}, ``{MeDiC GPGPU-Sim Patch -- GitHub Repository},''
  \url{https://github.com/CMU-SAFARI/MeDiC}.

\bibitem{memschedsim.github}
{SAFARI Research Group}, ``{MemSchedSim -- GitHub Repository},''
  \url{https://github.com/CMU-SAFARI/MemSchedSim}.

\bibitem{mosaic.github}
{SAFARI Research Group}, ``{Mosaic Simulator -- GitHub Repository},''
  \url{https://github.com/CMU-SAFARI/Mosaic}.

\bibitem{ramulator.github}
{SAFARI Research Group}, ``{Ramulator -- GitHub Repository},''
  \url{https://github.com/CMU-SAFARI/ramulator}.

\bibitem{safari.github}
{SAFARI Research Group}, ``{SAFARI Software Tools -- GitHub Repository},''
  \url{https://github.com/CMU-SAFARI/}.

\bibitem{strukov.nature08}
D.~B. Strukov, G.~S. Snider, D.~R. Stewart, and R.~S. Williams, ``{The Missing
  Memristor Found},'' \emph{Nature}, 2008.

\bibitem{subramanian.iccd14}
L.~Subramanian, D.~Lee, V.~Seshadri, H.~Rastogi, and O.~Mutlu, ``{The
  Blacklisting Memory Scheduler: Achieving High Performance and Fairness at Low
  Cost},'' in \emph{ICCD}, 2014.

\bibitem{subramanian.tpds16}
L.~Subramanian, D.~Lee, V.~Seshadri, H.~Rastogi, and O.~Mutlu, ``{BLISS:
  Balancing Performance, Fairness and Complexity in Memory Access
  Scheduling},'' \emph{TPDS}, 2016.

\bibitem{predictable-safarij}
L.~Subramanian, V.~Seshadri, Y.~Kim, B.~Jaiyen, and O.~Mutlu, ``{Predictable
  Performance and Fairness through Accurate Slowdown Estimation in Shared Main
  Memory Systems},'' \url{https://arxiv.org/abs/1805.05926}.

\bibitem{subramanian.hpca13}
L.~Subramanian, V.~Seshadri, Y.~Kim, B.~Jaiyen, and O.~Mutlu, ``{MISE:
  Providing Performance Predictability and Improving Fairness in Shared Main
  Memory Systems},'' in \emph{HPCA}, 2013.

\bibitem{subramanian.thesis15}
L.~Subramanian, ``{Providing High and Controllable Performance in Multicore
  Systems Through Shared Resource Management},'' Ph.D. dissertation, Carnegie
  Mellon Univ., 2015.

\bibitem{subramanian.micro15}
L.~Subramanian, V.~Seshadri, A.~Ghosh, S.~Khan, and O.~Mutlu, ``{The
  Application Slowdown Model: Quantifying and Controlling the Impact of
  Inter-Application Interference at Shared Caches and Main Memory},'' in
  \emph{MICRO}, 2015.

\bibitem{zorua-safarij}
N.~Vijaykumar, K.~Hsieh, G.~Pekhimenko, S.~Khan, A.~Shrestha, S.~Ghose, A.~Jog,
  P.~B. Gibbons, and O.~Mutlu, ``{Decoupling GPU Programming Models from
  Resource Management for Enhanced Programming Ease, Portability, and
  Performance},'' \url{https://arxiv.org/abs/1805.02498}.

\bibitem{vijaykumar.micro16}
N.~Vijaykumar, K.~Hsieh, G.~Pekhimenko, S.~Khan, A.~Shrestha, S.~Ghose, A.~Jog,
  P.~B. Gibbons, and O.~Mutlu, ``{Zorua: A Holistic Approach to Resource
  Virtualization in GPUs},'' in \emph{MICRO}, 2016.

\bibitem{wong.procieee12}
H.-S.~P. Wong, H.-Y. Lee, S.~Yu, Y.-S. Chen, Y.~Wu, P.-S. Chen, B.~Lee, F.~T.
  Chen, and M.-J. Tsai, ``{Metal-Oxide RRAM},'' \emph{Proc. IEEE}, 2012.

\bibitem{wong.procieee10}
H.-S.~P. Wong, S.~Raoux, S.~Kim, J.~Liang, J.~P. Reifenberg, B.~Rajendran,
  M.~Asheghi, and K.~E. Goodson, ``{Phase Change Memory},'' \emph{Proc. IEEE},
  2010.

\bibitem{yoon.taco14}
H.~Yoon, J.~Meza, N.~Muralimanohar, N.~P. Jouppi, and O.~Mutlu, ``{Efficient
  Data Mapping and Buffering Techniques for Multi-Level Cell Phase-Change
  Memories},'' \emph{TACO}, 2014.

\bibitem{rbla-safarij}
H.~Yoon, J.~Meza, R.~Ausavarungnirun, R.~A. Harding, and O.~Mutlu, ``{A Memory
  Controller with Row Buffer Locality Awareness for Hybrid Memory Systems},''
  \url{https://arxiv.org/abs/1804.11040}.

\bibitem{yoon.iccd12}
H.~Yoon, J.~Meza, R.~Ausavarungnirun, R.~A. Harding, and O.~Mutlu, ``{Row
  Buffer Locality Aware Caching Policies for Hybrid Memories},'' in
  \emph{ICCD}, 2012.

\bibitem{zhou.isca09}
P.~Zhou, B.~Zhao, J.~Yang, and Y.~Zhang, ``{A Durable and Energy Efficient Main
  Memory Using Phase Change Memory Technology},'' in \emph{ISCA}, 2009.

\end{thebibliography}
}

\end{document}